\documentclass[ notitlepage,letterpaper]{article}
\usepackage[ansinew]{inputenc} 
\usepackage[colorlinks=true,urlcolor=blue,linkcolor=blue,citecolor=blue]{hyperref} 
\usepackage{graphicx}
\usepackage{amsfonts}
\usepackage{amssymb}
\usepackage{color}
\usepackage{harvard}
\usepackage{cite}

\usepackage{amsthm}
\usepackage{amsmath}

\begin{document}

\title{Cracking in charged relativistic spheres}
\author{
\textbf{Guillermo A. Gonz\'{a}lez}\thanks{\texttt{guillego@uis.edu.co}}, \textbf{Anamar\'ia Navarro}\thanks{\texttt{anamaria.navarro@correo.uis.edu.co}} \\
\textit{Grupo de Investigaci\'on en Relatividad y Gravitaci\'on,} \\
\textit{ Escuela de F\'isica, Universidad Industrial de Santander,  }\\ 
\textit{A. A. 678, Bucaramanga 680002, Colombia}; \\
 and \textbf{Luis A. N\'{u}\~{n}ez}\thanks{\texttt{lnunez@uis.edu.co}} \\
 \textit{Grupo de Investigaci\'on en Relatividad y Gravitaci\'on,} \\
\textit{ Escuela de F\'isica, Universidad Industrial de Santander,  }\\ 
\textit{A. A. 678, Bucaramanga 680002, Colombia} and \\
\textit{Centro de F\'{\i}sica Fundamental, Departamento de F\'{\i}sica,} \\ 
\textit{Universidad de Los Andes, M\'{e}rida 5101, Venezuela.} 
}
\maketitle

\begin{abstract}
Using the concept of cracking, we have explored the influence of density fluctuations on isotropic and  anisotropic charged matter configurations in General Relativity with ``barotropic'' equations of state, $P = P(\rho)$  and $P_{\perp}= P_{\perp}(\rho)$ and a mass-charge relation $Q=Q(\rho)$.  We have refined the idea that density fluctuations affect physical variables and their gradients, i.e. the radial pressure and charge density gradients.  It is found that not only anisotropic charged models could present cracking (or overturning), but also isotropic charged matter configurations could be affected by density fluctuations. 
\end{abstract}


\section{Introduction}
Despite that there exist some consensus about the charge neutrality of astrophysical objects, the debate dedicated to the influence of the electric charge on the structure and evolution of relativistic self-gravitating systems starts almost with the theory of General Relativity. Following the early contributions of Rosseland and Eddington  \cite{Rosseland1924,Eddington1926} there is a long list of works concerning charge and relativistic compact objects (almost twenty thousands entries in scholar.google.com and more than six thousands in last five years). There has been a renewed interest in this subject emerging from the very ingenious theoretical proposals for mechanisms allowing for the appearance of electric charge in strange stars with quark matter \cite{MosqueraPennaPerez2003,Usov2004,MakHarko2004,UsovHarkoCheng2005,NegreirosEtal2009,NegreirosEtal2010,MannarelliEtal2014}  and the consequences on the compact object structure of particular charge distributions regarding concepts such as global and local neutrality (see \cite{PachonRuedaSanabria2006, RotondoEtal2011, RuedaEtal2012, BelvedereEtal2014}). Obviously, all the above models are worthless if they are unstable against fluctuations of their physical variables and efforts to study of stability of neutral and charged matter configurations are as old as the modelling itself. 

One of the most common strategies to study the stability of selfgravitating fluids is to follow the evolution in time of the perturbations of the physical variables. In General Relativity this methology was developed in 1964, for pascalian fluids, by S. Chandrasekhar \cite{Chandrasekhar1964a,Chandrasekhar1964b,Chandrasekhar1964c} and later it was extended to non-pascalian -i.e. anisotropic fluids having unequal radial and tangential pressures: $P_{r} \neq P_{\perp}$- by Hillebrandt and Steinmetz in 1976\cite{HillebrandtSteinmetz1976} and, more recently, by Dev and Gleiser\cite{DevGleiser2003}. 
Despite this effort, Herrera and coworkers \cite{Herrera1992,  DiPriscoEtal1994, DiPriscoHerreraVarela1997}, following a qualitatively different point of view, introduce the concept of cracking (or overturning) to describe the behaviour of fluid distributions just after its departure from equilibrium, when total non-vanishing radial forces of different signs appear within the system. They illustrated, under this approach, that some other possible additional pathologies of the fluid could be present in a matter distribution. Recently, \cite{GonzalezNavarroNunez2014, GonzalezNavarroNunez2015}, considering models described by ``barotropic'' equations of state $P=P(\rho)$ and $P_\perp = P_\perp(\rho)$,  extends this criterion to isotropic and anisotropic spheres under local perturbations of the density that affect not only the physical variables but also their gradient. This extension leads to an straight forward generalisation of the cracking criterium for anisotropic charged fluids which is developed in the present paper where we depart from the conjecture that the charge and the mass of a fluid are related by a ``barotropic equation of state'', $Q = Q(\rho)$. Following the same scheme developed in \cite{GonzalezNavarroNunez2014, GonzalezNavarroNunez2015} apply this idea to non constant (local) density fluctuations that can affect the gradient of pressure and charge density considering the local fluctuations, i.e. $\delta \rho = \delta \rho(r)$, represented by any function of compact support defined in a closed interval $\Delta r \ll 1$. 

This work is organised as follows. Section \ref{HydrostaticChargedSphere} will present the context $\&$ notation. In Section \ref{InstabCracking} we describe the concept of cracking for selfgravitating charged matter configurations presenting the effects of local density fluctuations on the force distributions within a matter distribution. Section \ref{Models}  illustrates these effects by some examples. Finally, some conclusions are displayed in Section \ref{Remarks}.

\section{Hydrostatic equilibrium of charged spheres}
\label{HydrostaticChargedSphere}

Let us consider an static spherically symmetric metric, described by the line-element
\begin{equation}
\mathrm{d}s^2 = \mathrm{e}^{2 \nu(r)}\,\mathrm{d}t^2-\mathrm{e}^{2\lambda(r)}\,\mathrm{d}r^2-r^2 \left(\mathrm{d}\theta ^2+
\sin^2\theta\,\mathrm{d}\phi^2\right) \; ,
\label{metricSpherical}
\end{equation}
with a matter component of the energy-momentum tensor that corresponds to an anisotropic fluid
\begin{equation}
{T}_{\mu \nu} = (\rho + P_{\perp}){{u}}_\mu{ {u}}_\nu - P_{\perp}{g}_{\mu \nu}  +
(P-P_{\perp}){{v}}_\mu {{v}}_\nu \label{tmunu_g} +  \frac{1}{4\pi} \left( F_{\mu \alpha} {F_{\nu}}^{\alpha} - \frac{1}{4}g_{\alpha \beta} F_{\alpha \beta} F^{\alpha \beta}  \right)  \; ,
\end{equation}
where $\rho$ is the energy density, $P$ the radial pressure, $P_{\perp}$ the tangential pressure,
\begin{equation}
{{u}}_\mu =  (\mathrm{e}^{\nu}, 0, 0, 0)  \,, \quad \mathrm{and} \quad 
{{v}}_\mu =  (0,-\mathrm{e}^{\lambda}, 0, 0) \; , \label{umu}
\end{equation}
are respectively the unitary velocity vector of a co-movil observer and a unitary normal vector to it. The electromagnetic field tensor $F_{\mu \nu}$ can be written in terms of the electromagnetic potential $A_{\mu}$,
\begin{equation}
F_{\mu \nu} = A_{\nu; \mu} - A_{\mu; \nu} \; , 
\end{equation}
therefore, the Maxwell electromagnetic field equations are 
\begin{equation}
\label{electromagnetic_field_equations}
\left( \sqrt{-g} F^{\mu \nu}\right)_{, \nu} = 4\pi \sqrt{-g} J^\mu  \; ,
\end{equation}
with $J^\mu = \sigma u^\mu$ the current four vector, and $\sigma $ the charge density.  

Since we want to consider an electric charged sphere, the potential vector is of the form
\begin{equation}
A_{\mu} = (A_0 , 0,0,0) \; ,
\end{equation}
then the only components of the electromagnetic tensor are $F_{01}$ and $-F_{10}$, and the Maxwell field equations (\ref{electromagnetic_field_equations}) can be written like
\begin{equation}
-\left( e^{-(\nu + \lambda)} r^2 F_{01} \right)_{, r} = 4\pi r^2 e^{\lambda} \sigma \; .
\end{equation}
Defining charge $Q(r)$ as
\begin{equation}
Q(r) = \int_0^r 4\pi \tau^2 \sigma e^{\lambda} \mathrm{d}\tau  \; ,
\end{equation}
we can express the components of the electromagnetic tensor $F_{01}$ as
\begin{equation}
F_{01} = -\frac{Q(r)}{r^2} e^{\lambda + \nu}  \; .
\end{equation}
and the non-null components of the energy-momentum tensor, $ T_{\mu \nu} $, as
\begin{align}
 & {T^{0}}_{0} = \rho + \frac{F_{01}^2}{8\pi} e^{-(\nu + \lambda)}  \; , \\
 & {T^{1}}_{1} = -P + \frac{F_{01}^2}{8\pi} e^{-(\nu + \lambda)}  \; , \\
 & {T^{2}}_{2} = {T^{3}}_{3} = -P_{\perp} - \frac{F_{01}^2}{8\pi} e^{-(\nu + \lambda)}   \; .
\end{align} 
Following \cite{Bekenstein1971}, we calculate $r-$component of the conservation equation, 
\begin{equation}
{T^{\mu}}_{1; \mu } = 0  \; ,
\end{equation}
and obtain
\begin{equation}
\label{R=0}
\frac{ d P }{d r}  - \frac{1}{4} \frac{Q }{ \pi r^4 } \frac{d Q}{d r} + \frac{ ( \rho + P )\left( 4 \pi r P + \frac{m}{r^2} - \frac{Q}{r^3} \right) }{(1- \frac{2m}{r} + \frac{Q^2}{r^2} )} - \frac{2(P_\perp - P)}{r} = 0  \; ,
\end{equation}
which is the generalization of the Tolman-Oppenheimer-Volkoff equation of hydrostatic equilibrium for a charged matter distribution. Since we are considering a static charged sphere, the exterior region is described by the Reissner-Norsdstrom metric
\begin{equation}
\mathrm{d}s^2 = \left(1 - \frac{2M}{r} + \frac{Q_0^2}{r^2} \right) \mathrm{d}t^2 - \frac{\mathrm{d}r^2 }{1 - \frac{2M}{r} + \frac{Q_0^2}{r^2} } -r^2 \left(\mathrm{d}\theta ^2 + \sin^2\theta\,\mathrm{d}\phi^2\right)  \; ,
\end{equation}
where $Q_0$ is the total charge and M is the total mass of the sphere. The inner region is matched to the Reissner-Norsdstrom metric through a null radial pressure hypersurface of radius $r_0$.

\section{Cracking of anisotropic charged spheres}
\label{InstabCracking}

In this section, following our previous work \cite{GonzalezNavarroNunez2014, GonzalezNavarroNunez2015}, we extend the concept of cracking for isotropic and anisotropic charged matter configurations. 

As we have stressed before this concept was introduced to describe the behaviour of fluid distributions just after they experiment a perturbation in any of its state variables and departs from equilibrium. Thus, the $r-$component of the conservation equation (\ref{R=0}) which represents radial forces may be different than zero and radial forces may appear on the system which could be represented by 
\begin{equation}
\label{R_charged}
\mathcal{R}  =\frac{ \mathrm{d} P }{\mathrm{d} r}  - \frac{1}{4} \frac{Q }{ \pi r^4 } \frac{\mathrm{d} Q}{\mathrm{d} r} + \frac{ ( \rho + P)\left( 4 \pi r P + \frac{m}{r^2} - \frac{Q}{r^3} \right) }{(1- \frac{2m}{r} + \frac{Q^2}{r^2} )} - \frac{2(P_\perp - P)}{r}  \; .
\end{equation}
 Herrera and collaborator \cite{Herrera1992,  DiPriscoEtal1994, DiPriscoHerreraVarela1997} had shown that there exist a relation between the force distribution and the expansion written as 
\begin{equation}
\mathcal{R} = - \frac{e^{\lambda}(\rho + P_{r})}{e^{\nu/2}r^{2}} \int_{0}^{a} \mathrm{d}\tilde{r} \ e^{\nu/2}\tilde{r}^{2}  \frac{\mathrm{d} \Theta}{\mathrm{d}s} \; ,
\label{Rintegral} 
\end{equation}
where $\Theta$ represents the expansion. Therefore if there is a change of sign on the force for some radial coordinate within the configuration, it affect also the expansion and  we say that cracking is produced.


Now let us consider an anisotropic charge sphere in hydrostatic equilibrium, described by equations of state of the form $P = P(\rho)$,  $P_\perp = P_\perp(\rho)$, and that hold a charge-mass relation of the form $Q = Q(\rho)$. Then, we will consider a local density perturbation 
\begin{equation}
\delta \rho = \delta\rho(r)  \; ,
\end{equation}
described by any function of compact support $\delta \rho = \delta \rho(r)$ defined in a closed interval $\Delta << 1$, which generate fluctuations in mass, radial and tangential pressures, and pressure gradient that can be represented as

\begin{align}
\label{expansions}
 & P(\rho + \delta \rho) \approx P(\rho) + \delta P = P(\rho) + \frac{\mathrm{d}P}{\mathrm{d}\rho}\delta \rho = P(\rho) +  v^2\delta \rho   \; , 
 \\  \nonumber \\   
  & P_\perp (\rho + \delta \rho)  \approx P_\perp(\rho) + \delta P_\perp = P_\perp (\rho) + \frac{\mathrm{d} P_\perp }{\mathrm{d} \rho} \delta \rho =  P_\perp (\rho)  + v_\perp^2 \delta \rho   \; , 
  \\ \nonumber  \\
  & P'(\rho + \delta \rho)  \approx P'(\rho) + \delta P' = P'(\rho) + \frac{\mathrm{d} P'}{\mathrm{d} \rho} \delta \rho \nonumber \\ 
 & \qquad \qquad \qquad \qquad  \qquad \quad  = P'(\rho) + \left[ (v^2)' + v^2 \frac{\rho''}{\rho'} \right] \delta \rho   \; , 
  \\ \nonumber \\
  & Q(\rho + \delta \rho)  \approx   Q(\rho) + \delta Q = Q(\rho) + \frac{\mathrm{d} Q}{\mathrm{d} \rho} \delta \rho = Q(\rho) + \frac{Q'}{\rho'} \delta \rho  \; ,
  \\ \nonumber \\
  & Q'(\rho + \delta \rho)  \approx Q'(\rho) + \frac{\partial Q'}{\partial \rho} \delta \rho = Q'(\rho) +  \frac{Q''}{\rho'}  \delta \rho   \; , 
  \\ \nonumber \\
  & m(\rho + \delta \rho)  \approx m(\rho) + \delta(m) = m(\rho) + \frac{\partial m}{\partial \rho} \delta \rho = m(\rho) + \frac{4\pi r^2 \rho}{\rho'} \delta \rho   \; .
\end{align}
Where primes denote radial differentiation while
\begin{equation}
v^2 =  \frac{\mathrm{d}P}{\mathrm{d}\rho}  \; , \quad v^2_\perp = \frac{\mathrm{d}P_\perp}{\mathrm{d}\rho}  \; , \quad m = 4\pi \int^r_0 \rho(\bar{r})\bar{r}^2 \mathrm{d}\bar{r}  \; , 
\end{equation}
are the radial sound speed, tangential sound speed and mass, respectively. 

Now, we can, formally, expand equation (\ref{R_charged}), and obtain the radial force that appears due to the perturbation of density
\begin{equation}
\label{R_plus_dR}
 \mathcal{R}  \approx  \mathcal{R} _0( \rho, m , P , P', P_\perp, Q, Q' )  +   \delta \mathcal{R}     \; ,
\end{equation}
with
\begin{equation}
 \delta \mathcal{R}   = \frac{\partial \mathcal{R}}{\partial \rho} \delta \rho + \frac{\partial \mathcal{R}}{\partial m} \delta m + \frac{\partial \mathcal{R}}{\partial P} \delta P 
 + \frac{\partial \mathcal{R}}{\partial P'} \delta P' + \frac{\partial \mathcal{R}}{\partial P_\perp} \delta P_\perp + \frac{\partial \mathcal{R}}{\partial Q}\delta Q + \frac{\partial \mathcal{R}}{\partial Q'}\delta Q'  \; ,
\end{equation}
and
\begin{equation}
\mathcal{R} _0( \rho, m , P, P', P_\perp, Q, Q' )  = 0  \; ,
\end{equation}
since the system is initially in equilibrium. 
By using (\ref{expansions}) it can be shown that
\begin{eqnarray}
  \delta \mathcal{R} = \delta \rho \left[ \frac{\partial \mathcal{R}}{\partial \rho} +
                       \frac{\partial \mathcal{R}}{\partial m} \left( \frac{4\pi r^2 \rho}{\rho'}\right)  + \frac{\partial \mathcal{R}}{\partial P} v^2  + \frac{\partial \mathcal{R}}{\partial P'} \left( (v^2)'+ v^2 \frac{\rho''}{\rho'}  \right)  \right.  \qquad  \nonumber \\
 \qquad \qquad  \qquad  \qquad \qquad  \qquad    \left. + \frac{\partial \mathcal{R}}{\partial P_\perp} v_\perp^2 + \frac{\partial \mathcal{R}}{\partial Q} \left( \frac{Q'}{\rho'}\right) + \frac{\partial \mathcal{R}}{\partial Q'} \left( \frac{Q''}{\rho'}  \right) \right]
\end{eqnarray}
where
\begin{align}
& \frac{\partial \mathcal{R}}{\partial \rho} = \frac{  4 \pi r^4 P + m r  - Q^2 }{r(r^2 - 2m r + Q^2)}  \; , \qquad
 \frac{\partial \mathcal{R}}{\partial m} = \frac{ (\rho + P)( r^2 - Q^2 + 8\pi r^4 P )  }{\left( r^2 - 2mr + Q^2\right)^2 }  \; ,  \\ \nonumber \\
&  \frac{ \partial \mathcal{R} }{ \partial Q } = -\frac{1}{4\pi r^4} \frac{\mathrm{d}Q}{\mathrm{d}r} - \frac{2\left(\rho + P \right)\left( 4\pi r^4 P + mr - Q^2 + 1 \right)Q}{r\left(r^2 - 2 mr + Q^2\right)} \; ,  \; , \\ \nonumber \\
 & \frac{\partial \mathcal{R}}{\partial P} = \frac{ 8\pi r^4 P - 3 m r  + Q^2 + 4 \pi r^4 \rho + 2 r^2 }{r(r^2 - 2 m r + Q^2)}  \; ,  \\ \nonumber \\
& \frac{\partial \mathcal{R}}{\partial Q'} = -\frac{1}{4\pi r^4} Q  \; , \qquad
 \frac{\partial \mathcal{R} }{\partial P_\perp} = -\frac{2}{r} \; , \qquad \frac{\partial \mathcal{R}}{\partial P'} = 1 
 \end{align}

As in \cite{ GonzalezNavarroNunez2014, GonzalezNavarroNunez2015}, if $\delta \mathcal{R}$ does not change its sign we will consider that the configuration is potentially stable \cite{DiPriscoHerreraVarela1997 , AbreuHernandezNunez2007b}. Since this time there are terms of the charge, and the charge gradient, is not simple to identify which terms are the ones that may contribute to the change of sign of the radial force; then the whole expression must be taken into consideration. In the next section we will analyse this term in some models in order to identify instabilities under this criterion, studying how the terms $\tilde{R}_i$ determine the outcome of the perturbation
\begin{equation}
\label{tilde_R}
\tilde{R} = \frac{\delta R}{\delta \rho} =  \tilde{R}_1 + \tilde{R}_2 + \tilde{R}_3 + \tilde{R}_4 + \tilde{R}_5 + \tilde{R}_7 + \tilde{R}_8   \; ,
\end{equation}
\begin{align}
 & \tilde{R}_1 =  \frac{\partial \mathcal{R}}{\partial \rho}  \; , \quad
 \tilde{R}_2 =  \frac{\partial \mathcal{R}}{\partial m} \frac{4\pi r^2 \rho}{\rho'}  \; , \quad
 \tilde{R}_3 =  \frac{\partial \mathcal{R}}{\partial P} v^2  \; , \qquad \tilde{R}_5  = \frac{\partial \mathcal{R}}{\partial P_\perp} v_\perp^2  \; ,
  \label{R1234} \\ \nonumber \\
 & \tilde{R}_4 = \frac{\partial \mathcal{R}}{\partial P'}  \left[ (v^2)' + v^2\frac{\rho''}{\rho'} \right]  \; , \qquad   \tilde{R}_6 = \frac{\partial \mathcal{R}}{\partial Q} \frac{Q'}{\rho'} \; , \quad
  \tilde{R}_7 = \frac{\partial \mathcal{R}}{\partial Q'} \frac{Q''}{\rho'} \label{R567}   \; .
\end{align}

\section{Analysis of charged models}
\label{Models}
In this section we illustrate, through examples, the three possible cases -cracking, overturning and no cracking at all- for isotropic and anisotropic fluid relativistic spheres.  

\subsection{Model 1: Isotropic charged sphere}
We analyse an isotropic charged sphere model proposed in \cite{SinghSinghHelmi1993}, equations (3.17) and
(3.18), described by the density, pressure, and charge distribution:
\begin{align}
 \tilde{\rho} &= \rho r_0^2 = \frac{ 5\tilde{r}^2 - 3 - 3\alpha_4 q^2 + 3\mu \alpha_4^4 - 3 \alpha_4^2 q^2}{8\pi \alpha_4^2} + \frac{q^2}{4\pi \tilde{r}^2} \, ,   \\ \nonumber \\
 P  &= \tilde{P} r_0^2  = \frac{2\sqrt{\alpha_4^2 + (1+\alpha_4 q^2 - \mu \alpha_4^2 + \alpha_4^2 q^2)\tilde{r}^2 - \tilde{r}^4 - \alpha_4^2 q^2 } }{8\pi \alpha_4^2} \, , \nonumber \\ 
&  \qquad \quad + \frac{ 1 + \alpha_4 q^2 (q^2 - \mu) \alpha_4^2 - \tilde{r}^2}{8\pi \alpha_4^2} + \frac{q^2}{8\pi \tilde{r}^2} \, , \\ \nonumber \\
 \tilde{Q} &= Q r_0^2 = q \tilde{r} \, , \\ \nonumber \\
 \tilde{P}_\perp &= \tilde{P} \, , 
\end{align}
with
\begin{align}
 \alpha_4 &= \alpha_3 - \frac{1}{3}\frac{(12\mu-12-q^4)(\mu^2 - 4q^2\mu + 4q^4)}{(\mu-2q^2)(\alpha_2(\mu - 2 q^2)^2(\mu^2-4q^2\mu+4q^4)^2)^{1/3}}  \nonumber \\
  &  \qquad \qquad \qquad \qquad \qquad  \qquad \qquad \qquad\qquad \qquad  + \frac{2}{3}\frac{q^2}{(\mu-2q^2)}  \, , \\ \nonumber \\
 \alpha_3 &= \frac{1}{3}\frac{\left(\alpha_2 \left(\mu-2q^2\right)^2 \left( \mu^2 - 4q^2\mu + 4q^4 \right)^2 \right)^{1/3}}{\left(\mu-2q^2\right)\left(\mu^2-4q^2\mu + 4 q^4\right)} \, , \\ \nonumber \\
 \alpha_2 &= 18 q^2 \mu + 36q^2 - q^6 - 108q^4 + 6\sqrt{3} \alpha_1  \, , \\ \nonumber \\
 \alpha_1 &= \left[ 16\mu^3 (- 48 - q^4)\mu^2 + (48 + 20 q^4 - 36q^6 )\mu -16  \right.  \nonumber \\ 
  &  \qquad \qquad \qquad \qquad \qquad \qquad \left.  + 2 q^10 + 8 q^4 + 107 q^8 - 72 q^6  \right]^{1/2} \, . 
\end{align}

\begin{figure}[h]
\begin{center}
\includegraphics[width = 0.45\textwidth]{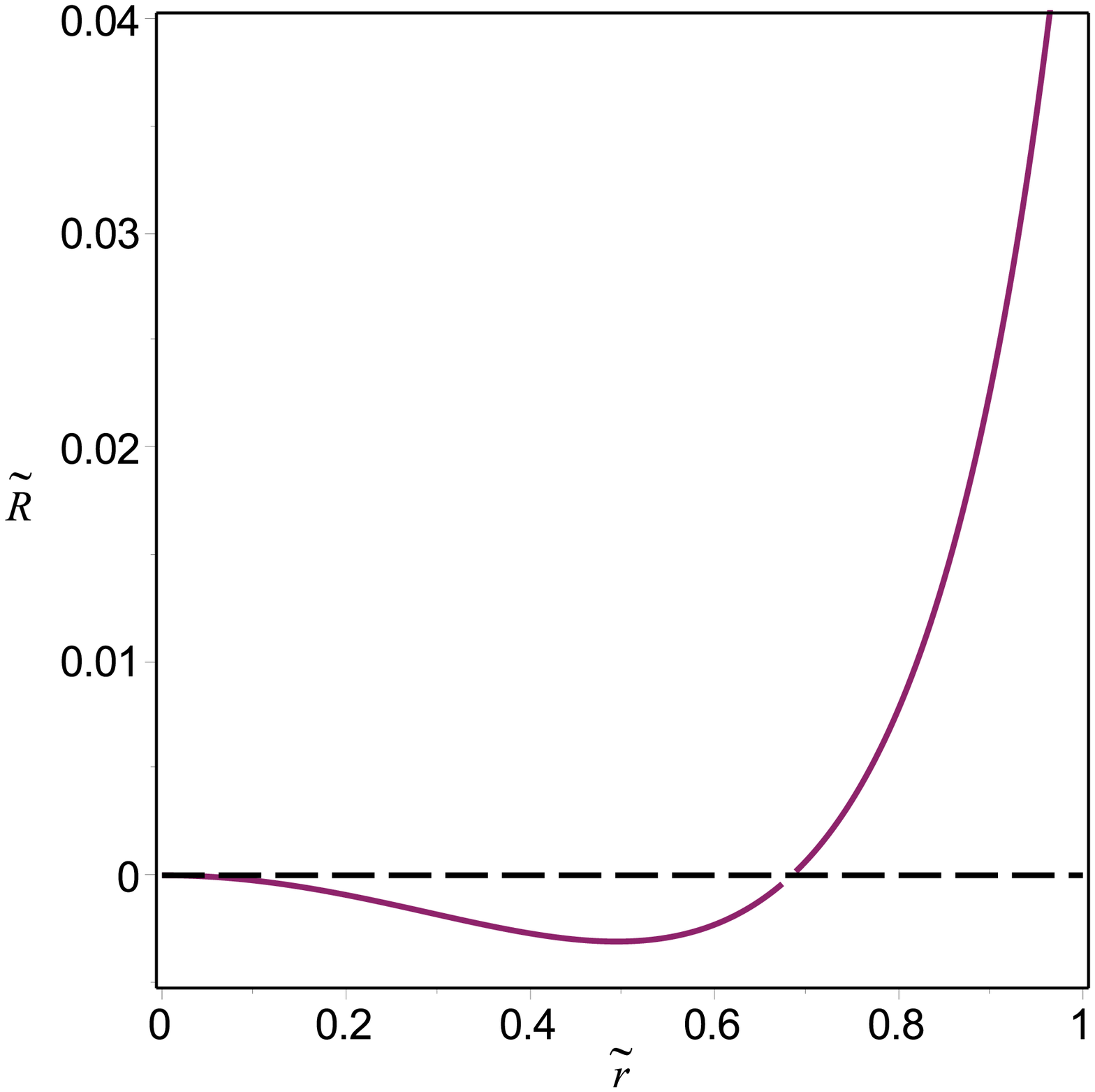} \includegraphics[width = 0.45\textwidth]{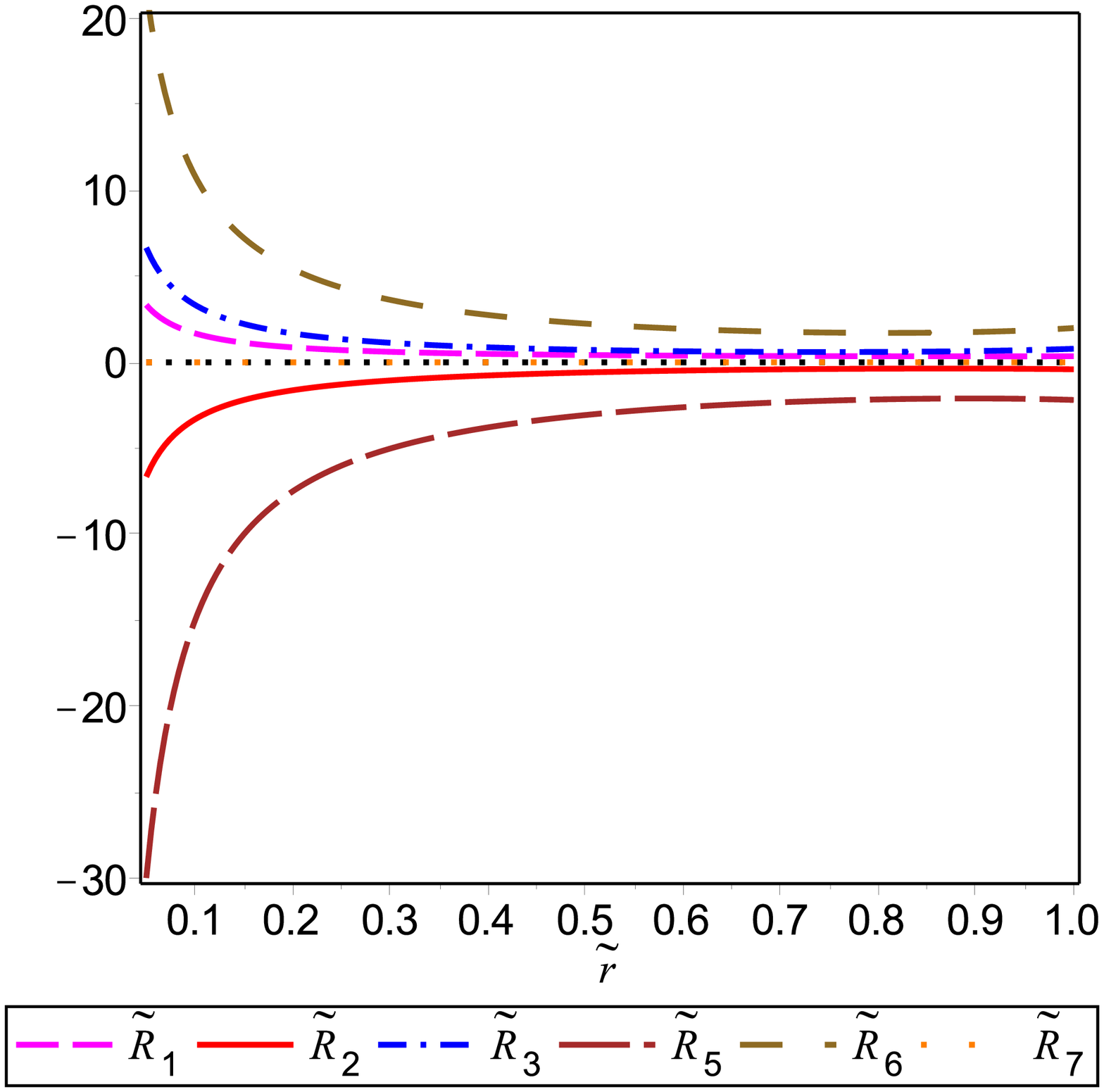}
\end{center} 
\caption{On the left, the force distribution, $\widetilde{\mathcal{R}}=\delta \mathcal{R} / \delta \rho$ for the isotropic charged sphere proposed in \cite{SinghSinghHelmi1993}, using the parameters $\mu = 0.1 $ and $ q = 0.1$.  Since the force changes sign around $\tilde{r}=0.7$, we say that the configuration could be potentially unstable under the perturbation of density $\delta \rho$  the region near that coordinate.  On the right  plate, all the terms that compose the force distribution $\widetilde{\mathcal{R}}_i$ are sketched . \label{Graficas_caso4}}
\end{figure}

Now, substituting the expressions for the density, charge and pressures, using the parameters $\mu =0.1$ and $q=0.1$. In figure \ref{Graficas_caso4} we plot $\tilde{R}$, of equation (\ref{tilde_R}), and the terms $\tilde{R}_i$ that compose it separately, equations (\ref{R1234}) to (\ref{R567}).  This model exhibits cracking since $\tilde{R} $ changes sign around $\tilde{r}=0.7$, this indicates that around that region the model could be potentially unstable under perturbations of the density. 

\subsection{Model 2: Anisotropic charged sphere I.}
Now, we analyse an anisotropic charged model proposed in \cite{EsculpiAloma2010}, described by
\begin{align}
 \tilde{\rho} &= \rho r_0^2 = \frac{1}{8} \frac{n + \tilde{r}^2 }{\pi \left( 1 + 2 n - C \right)\tilde{r}^2} \; , \\ \nonumber \\
 \tilde{P} &= P r_0^2 = \frac{1}{8} \frac{1 + \tilde{r}^2 }{\pi \left( 1 + 2 n - C \right)\tilde{r}^2} \,, \\ \nonumber \\
  \tilde{P}_\perp &= P_\perp r_0^2 = C \tilde{P}  \,, \\ \nonumber \\
  \tilde{Q} &= Q r_0 = \sqrt{\frac{\tilde{r}^2\left(1 + n - 2C - \tilde{r}^2(1-C) \right)}{2+4n - 2C}} \,, 
\end{align}
This model represents a matter distribution, admitting a one parameter group of conformal motions with a linear equation of state.

\begin{figure}[h]
\begin{center}
\includegraphics[width = 0.45\textwidth]{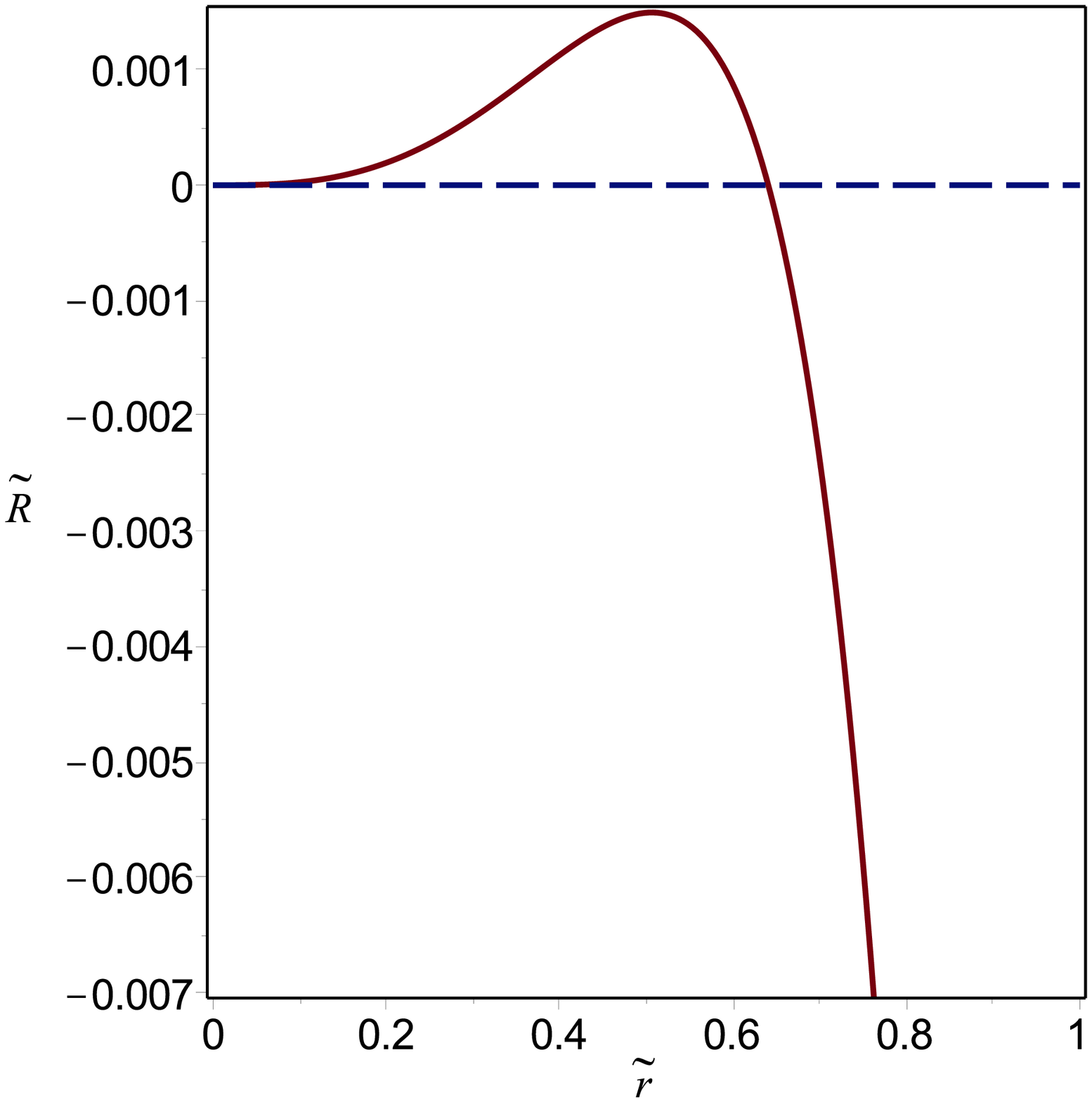} \includegraphics[width = 0.45\textwidth]{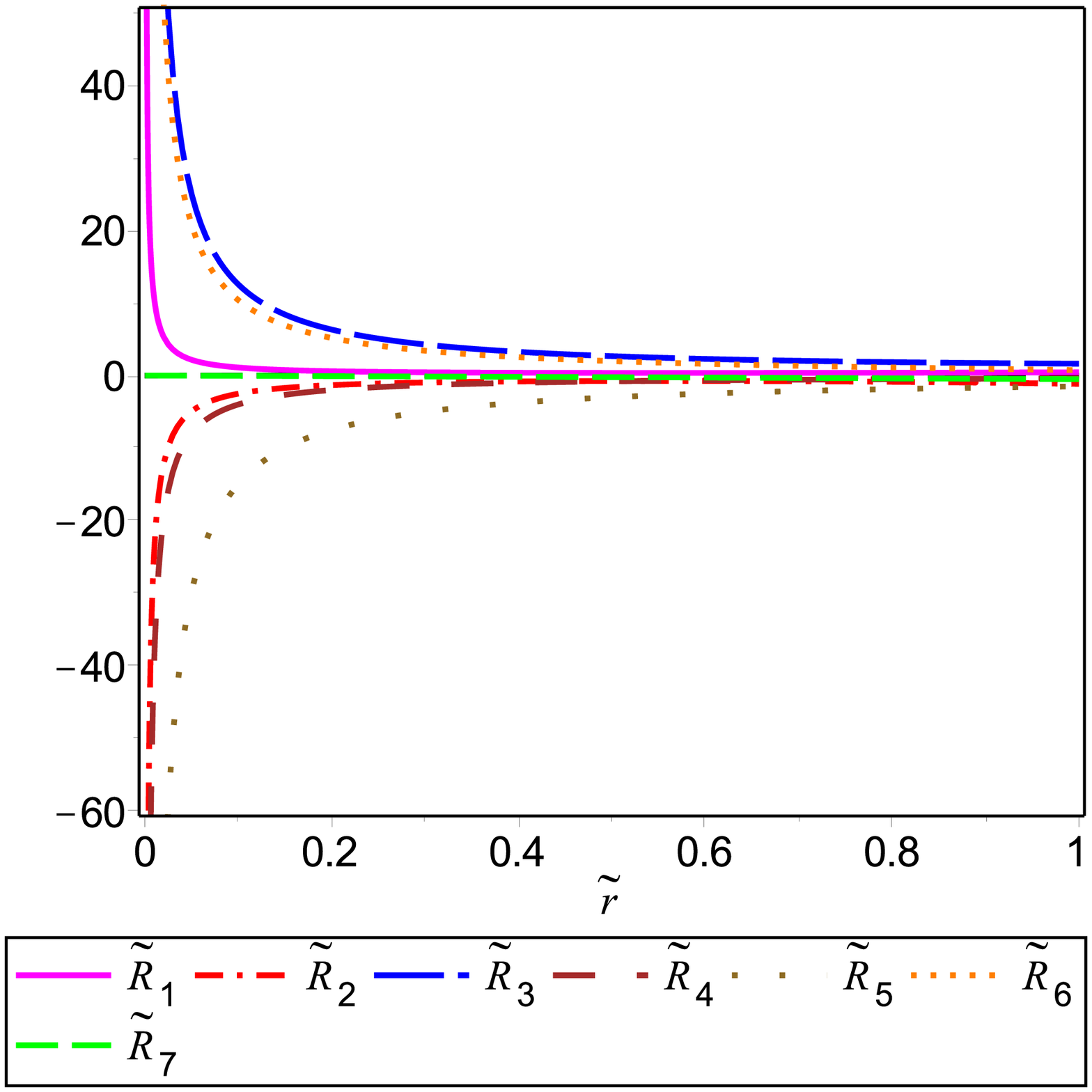}
\end{center} 
\caption{On the left, the force distribution, $\widetilde{\mathcal{R}}=\delta \mathcal{R} / \delta \rho$ for an anisotropic charged sphere proposed in\cite{EsculpiAloma2010}, using the parameters $C = (n-1)/2 - 0.1 $ and $ n = 2$. Since the force changes sign around $\tilde{r}=0.6$, we say that the configuration could be potentially unstable under the perturbation of density $\delta \rho$ around the region near that radial coordinate.  Again, on the right plate all the terms that compose the force distribution $\widetilde{\mathcal{R}}_i$. \label{Graficas_caso5}}
\end{figure}

Now, substituting the expressions for the density, charge and pressures, using the parameters $C = (n-1)/2 - 0.1 $ and $ n = 2$. In figure \ref{Graficas_caso5} we plot $\tilde{R}$, of equation (\ref{tilde_R}), and the terms $\tilde{R}_i$ that compose it separately, equations (\ref{R1234}) to (\ref{R567}). Now, as well as the previous example, this model is potentially unstable since $\tilde{R} $ changes its sign around $\tilde{r}=0.6$. 

\subsection{Model 3: Anisotropic charged sphere II}
\begin{figure}[h]
\begin{center}
\includegraphics[width = 0.45\textwidth]{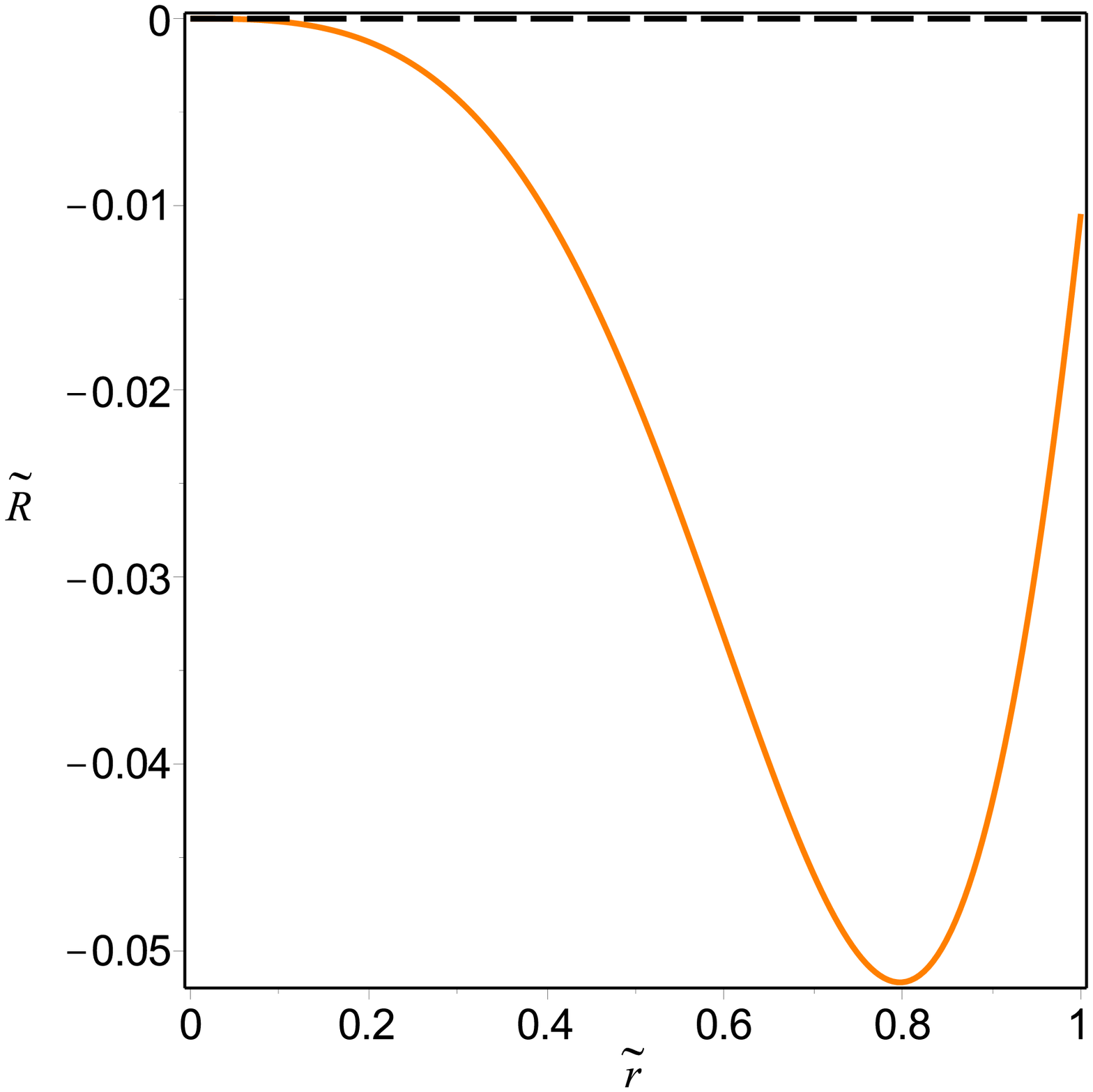} \includegraphics[height = 0.45\textwidth]{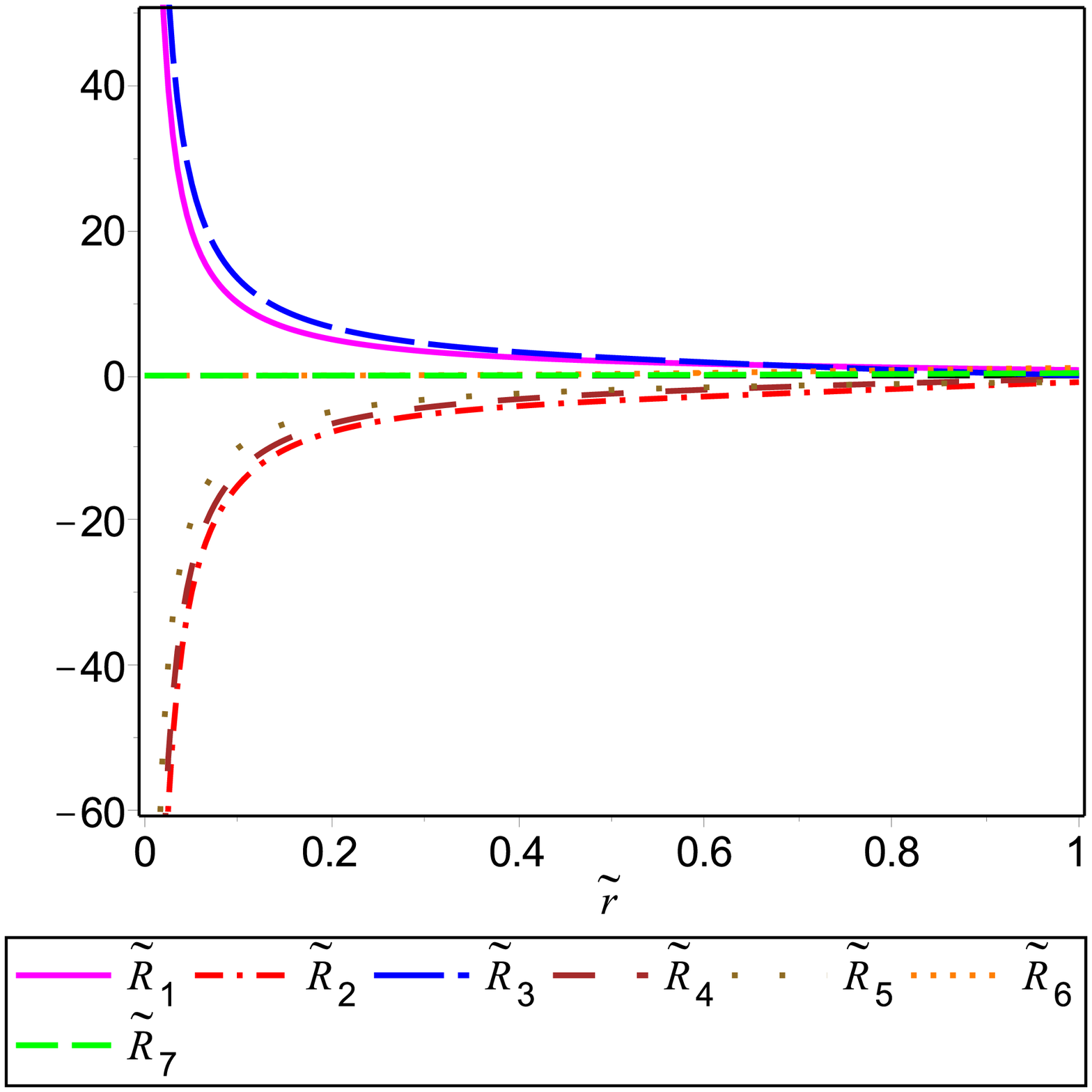}
\end{center} 
\caption{On the left plate, the force distribution, $\widetilde{\mathcal{R}}=\delta \mathcal{R} / \delta \rho$ for an anisotropic charged sphere \cite{HerreraPonce1985}, using the parameters $C_1 = 0.4$ and $\alpha = 1$ . Since the force does not change sign, no cracking occurs and the configuration is stable under the perturbation of density $\delta \rho$.  On the right, the terms that compose the force distribution $\widetilde{\mathcal{R}}_i$. \label{Graficas_caso3}} 
\end{figure}

Now, let us consider an anisotropic charged model proposed also in \cite{HerreraPonce1985}, equations (125), (126) and (127), described by the dimensionless equations
\begin{align}
 \tilde{\rho} &= \rho r_0^2 = \frac{\left( 3 C_1 - 1 \right) }{8\pi \tilde{r}^2} \left[  \frac{(1 - C_1) }{\tilde{r}^2(3C_1 - 1) } + 1 + \alpha^2 \left( 1 - \frac{3\tilde{r}^2}{2} \right) \right]  \; , \\ \nonumber \\
 \tilde{P} &= P r_0^2 = \frac{\left( 3C_1-1\right)}{8\pi \tilde{r}^2} \left(1 - \tilde{r}^2 \right) \left(1 - \alpha^2 \tilde{r}^2 \right)  \; , \\ \nonumber \\
 \tilde{P}_\perp &= P_\perp r_0^2 = \frac{\left(1 - 2 C_1 \right)}{8\pi \tilde{r}^2 } + \tilde{P}  \; , \\ \nonumber \\
 \tilde{Q} &= \frac{Q}{r_0} = \frac{\sqrt{3\alpha C_1 - \alpha} \tilde{r}^3}{2} \; .
\end{align}
Now, substituting the expressions for the density, charge and pressures, using the parameters $C_1 = 0.4$ and $\alpha = 1$, in figure \ref{Graficas_caso3} we plot $\tilde{R}$, of equation (\ref{tilde_R}), and the terms $\tilde{R}_i$ that compose it separately, equations (\ref{R1234}) to (\ref{R567}). As well as the previous example, this model has cracking instability since $\tilde{\mathcal{R}}$ changes sign around $r\approx 0.997$. 

\section{Remarks and conclusions}
\label{Remarks}
We have shown that isotropic and anisotropic charged matter configurations can present cracking and overturning when non constant (local) density fluctuations are considered and affect the state variables and their gradient. As we have pointed out, the density perturbations we assume are local, $\delta \rho = \delta \rho(r)$ -represented by any function of compact support defined in a closed interval $\Delta r \ll 1$- within a fluid configuration governed by barothropic equations of state: $P = P(\rho)$, $P_\perp = P_\perp(\rho)$ and a mass-charge relation $Q = Q(\rho)$. 

 The idea of cracking was originally conceived by Herrera\cite{Herrera1992} to describe the behaviour of fluid distributions just  after their departure from equilibrium: fluid elements, at both sides of the cracking point are accelerated with respect to each other by -independent and simultaneous- perturbations in energy density and  anisotropy \cite{DiPriscoEtal1994,DiPriscoHerreraVarela1997}. In their approach they considered independent and simultaneous perturbations in density and anisotropy that lead cracking (or overturning) within the configuration. Later, Abreu, Hern\'andez and N\'{u}\~{n}ez \cite{AbreuHernandezNunez2007b} shown how constant (global) density perturbations could generate cracking on anisotropic relativistic fluids.  In this study, constant density fluctuations affect  mass,  radial and  tangential pressure, but leave unperturbed the pressure gradient; again only anisotropic distributions can exhibit cracking or overturning. This idea was extended to the charged case in \cite{ManjarresNunezPercoco2008} where some preliminaries results were presented. In a recent work \cite{GonzalezNavarroNunez2014, GonzalezNavarroNunez2015}, we extend this criterium -of density perturbation affecting the state variables and their gradient- and showed that cracking (or overturning) could also be present for isotropic and anisotropic spheres. This extension leads to an straight forward generalisation of the cracking criterium for isotropic/anisotropic charged fluids developed in the present paper.

It is worth to be mentioned that, the concept of cracking illustrate that some pathology could be present for a particular equation of state and refers only  to the tendency of the configuration to split (or to compress) at a particular point within the distribution but not to collapse or to expand. The cracking, overturning, expansion or collapse, has to be established from the integration of the full set of Einstein  equations, but it should be clear that the occurrence of these phenomena could drastically alter the subsequent evolution of the system. If within a particular configuration no cracking (or overturning) is present, we could identify it as  \textit{potentially} stable (not absolutely stable), because other types of perturbations could lead to its expansions or collapse. Within a relativistic matter configuration, in principle, it is not clear which of the two scenarios is more likely to occur: the  simultaneous two-perturbation original scenario of Herrera and coworkers or the only density-driven framework but surely both generate instabilities on relativistic matter configurations that should be evaluated.

\section*{Acknowledgements}
Two of us, GAG and LAN gratefully acknowledge the financial support from Vicerrector\'{\i}a de Investigaci\'on y Extensi\'on, Universidad Industrial de Santander, while AN wants to thank the opportunity to be part of the program of Jóvenes Investigadores of COLCIENCIAS, Colombia.


\begin{thebibliography}{10}

\bibitem{Rosseland1924}
S.~{Rosseland}.
\newblock {Electrical state of a star}.
\newblock {\em Monthly Notices of the Royal Astronomical Society}, 84:720--728,
  June 1924.

\bibitem{Eddington1926}
A.~Eddington.
\newblock {\em The Internal Constitution of Star}.
\newblock Cambridge University Press, 1926.

\bibitem{MosqueraPennaPerez2003}
H.~J. {Mosquera-Cuesta}, A.~{Penna-Firme}, and A.~P\'erez-Lorenzana.
\newblock Charge asymmetry in the brane world and formation of charged black
  holes.
\newblock {\em Phys. Rev. D}, 67(8):087702, 2003.

\bibitem{Usov2004}
V.~V. Usov.
\newblock {Electric fields at the quark surface of strange stars in the
  color-flavor locked phase}.
\newblock {\em Phys. Rev.}, D70:067301, 2004.

\bibitem{MakHarko2004}
M.~K. Mak and T.~Harko.
\newblock {Quark stars admitting a one-parameter group of conformal motions}.
\newblock {\em Int. J. Mod. Phys.}, D13:149--156, 2004.

\bibitem{UsovHarkoCheng2005}
VV~Usov, Tiberiu Harko, and KS~Cheng.
\newblock Structure of the electrospheres of bare strange stars.
\newblock {\em The Astrophysical Journal}, 620(2):915, 2005.

\bibitem{NegreirosEtal2009}
Rodrigo~Pican{\c{c}}o Negreiros, Fridolin Weber, Manuel Malheiro, and Vladimir
  Usov.
\newblock Electrically charged strange quark stars.
\newblock {\em Physical Review D}, 80(8):083006, 2009.

\bibitem{NegreirosEtal2010}
Rodrigo~Pican{\c{c}}o Negreiros, Igor~N Mishustin, Stefan Schramm, and Fridolin
  Weber.
\newblock Properties of bare strange stars associated with surface electric
  fields.
\newblock {\em Physical Review D}, 82(10):103010, 2010.

\bibitem{MannarelliEtal2014}
Massimo Mannarelli, Giulia Pagliaroli, Alessandro Parisi, and Luigi Pilo.
\newblock Electromagnetic signals from bare strange stars.
\newblock {\em Physical Review D}, 89(10):103014, 2014.

\bibitem{PachonRuedaSanabria2006}
L.A. Pach{\'o}n, J.A. Rueda, and J.D. Sanabria-G{\'o}mez.
\newblock {Realistic exact solution for the exterior field of a rotating
  neutron star}.
\newblock {\em Physical Review D}, 73(10):104038, 2006.

\bibitem{RotondoEtal2011}
M~Rotondo, Jorge~A Rueda, Remo Ruffini, and S-S Xue.
\newblock Relativistic thomas-fermi treatment of compressed atoms and
  compressed nuclear matter cores of stellar dimensions.
\newblock {\em Physical Review C}, 83(4):045805, 2011.

\bibitem{RuedaEtal2012}
J.~A. {Rueda}, M.~{Rotondo}, R.~{Ruffini}, and S.-s. {Xue}.
\newblock {A New Family of Neutron Star Models: Global Neutrality versus Local
  Neutrality}.
\newblock In {\em Twelfth Marcel Grossmann Meeting on General Relativity}, page
  1039, 2012.

\bibitem{BelvedereEtal2014}
Riccardo Belvedere, Kuantay Boshkayev, Jorge~A Rueda, and Remo Ruffini.
\newblock Uniformly rotating neutron stars in the global and local charge
  neutrality cases.
\newblock {\em Nuclear Physics A}, 921:33--59, 2014.

\bibitem{Chandrasekhar1964a}
S.~Chandrasekhar.
\newblock {Dynamical Instability of Gaseous Masses Approaching the
  Schwarzschild Limit in General Relativity}.
\newblock {\em Physical Review Letters}, 12(4):114--116, 1964.

\bibitem{Chandrasekhar1964b}
S.~{Chandrasekhar}.
\newblock {The Dynamical Instability of Gaseous Masses Approaching the
  Schwarzschild Limit in General Relativity.}
\newblock {\em Astrophys. J.}, 140:417, August 1964.

\bibitem{Chandrasekhar1964c}
S~Chandrasekhar.
\newblock A general variational principle governing the radial and the
  non-radial oscillations of gaseous masses.
\newblock {\em The Astrophysical Journal}, 139:664, 1964.

\bibitem{HillebrandtSteinmetz1976}
W.~Hillebrandt and K.~O. Steinmetz.
\newblock {Anisotropic neutron star models-Stability against radial and
  nonradial pulsations}.
\newblock {\em Astronomy and Astrophysics}, 53(2), 1976.

\bibitem{DevGleiser2003}
K.~Dev and M.~Gleiser.
\newblock Anisotropic stars ii: Stability.
\newblock {\em Gen. Relativ. Gravitation}, 35(8):1435--1457, 2003.

\bibitem{Herrera1992}
L.~Herrera.
\newblock Cracking of self-gravitating compact objects.
\newblock {\em Phys. Lett. A}, 165(206-210), 1992.

\bibitem{DiPriscoEtal1994}
A.~Di~Prisco, E.~Fuenmayor, L.~Herrera, and V.~Varela.
\newblock Tidal forces and fragmentation of self-gravitating compact objects.
\newblock {\em Phys. Lett. A}, 195:23 -- 26, 1994.

\bibitem{DiPriscoHerreraVarela1997}
A.~Di~Prisco, L.~Herrera, and V.~Varela.
\newblock Cracking of homogeneous self-gravitating compact objects induced by
  fluctuations of local anisotropy.
\newblock {\em Gen. Relativ. Gravitation}, 29(10):1239--1256, 1997.

\bibitem{GonzalezNavarroNunez2014}
Guillermo~A Gonzalez, Anamaria Navarro, and Luis~A Nunez.
\newblock Cracking and instability of isotropic and anisotropic relativistic
  spheres.
\newblock {\em arXiv preprint arXiv:1410.7733}, 2014.

\bibitem{GonzalezNavarroNunez2015}
Guillermo~A Gonz{\'a}lez, Anamar{\'{\i}}a Navarro, and Luis~A. N\'u{\~n}ez.
\newblock Cracking of anisotropic spheres in general relativity revisited.
\newblock {\em Journal of Physics: Conference Series}, 600(1):012014, 2015.

\bibitem{Bekenstein1971}
J.~D. Bekenstein.
\newblock Hydrostatic equilibrium and gravitational collapse of relativistic
  charged fluid balls.
\newblock {\em Phys. Rev. D}, 4(8):2185--2190, Oct 1971.

\bibitem{AbreuHernandezNunez2007b}
H.~Abreu, H.~Hern{\'a}ndez, and L.~A. N{\'u}\~nez.
\newblock Sound speeds, cracking and stability of self-gravitating anisotropic
  compact objects.
\newblock {\em Class. Quant. Grav.}, 24:4631--4646, 2007.

\bibitem{SinghSinghHelmi1993}
T.~Singh, G.P. Singh, and A.M. Helmi.
\newblock {Solutions of Einstein's field equations for charged static fluid
  spheres}.
\newblock {\em Astrophys Space Sci}, 199(1):113--123, January 1993.

\bibitem{EsculpiAloma2010}
M.~Esculpi and E.~Alom\'a.
\newblock {Conformal anisotropic relativistic charged fluid spheres with a
  linear equation of state }.
\newblock {\em Eur. Phys. J. C}, 67(3-4):521--532, June 2010.

\bibitem{HerreraPonce1985}
L.~Herrera and J.~Ponce~de Le\'on.
\newblock {Isotropic and anisotropic charged spheres admitting a one-parameter
  group of conformal motions}.
\newblock {\em J. Math. Phys.}, 26:2302--2307, September 1985.

\bibitem{ManjarresNunezPercoco2008}
J.~Manjarr\'es, L.A. N{\'u}\~nez, and U.~Percoco.
\newblock Perturbaciones de carga en objetos compactos.
\newblock {\em Revista Integraci{\'o}n}, 25(2):147--152, Nov 2008.

\end{thebibliography}

\end{document}